# The Nambu-Jona-Lasinio model with staggered fermions*

A. Ali Khan [a]  M. Göckeler[a]  R. Horsley[a,b]  P.E.L. Rakow[c]  G. Schierholz[b,a]  and H. Stüben[c]

[a]Höchstleistungsrechenzentrum HLRZ, c/o Forschungszentrum Jülich, D-52425 Jülich, Germany

[b]Deutsches Elektronen-Synchrotron DESY, Notkestraße 85, D-22603 Hamburg, Germany

[c]Institut für Theoretische Physik, Freie Universität Berlin, Arnimallee 14, D-14195 Berlin, Germany

We investigate the neighbourhood of the chiral phase transition in a lattice Nambu–Jona-Lasinio model, using both Monte Carlo methods and lattice Schwinger-Dyson equations.

## 1. INTRODUCTION

We report the results of an investigation into a lattice Nambu-Jona-Lasinio model. For further details and references the reader is referred to [1, 2].

The Nambu–Jona-Lasinio model is a purely fermionic model. The interaction of the particles is given by a chirally invariant four fermion interaction [3]

$$S = \int d^4x \left\{ \bar{\psi}(x)\gamma_\mu \partial_\mu \psi(x) - g_0 \left[ \left( \bar{\psi}(x)\psi(x) \right)^2 - \left( \bar{\psi}(x)\gamma_5 \psi(x) \right)^2 \right] \right\} . (1)$$

Our investigation was motivated by the fact that four fermion interactions have become popular again. They appear for example in technicolour models, top-quark condensate models [4] and extensions of QED.

In their original Hartree-Fock calculation [3] Nambu and Jona-Lasinio found that for large values of the coupling $g_0$ the fermion mass $\mu_R$ is generated dynamically and that there exist two bound states of a fermion and an antifermion. There is scalar state with mass $2\mu_R$ and a massless pseudoscalar state which is the Goldstone-boson associated with the spontaneous breakdown of chiral symmetry.

In order to achieve an understanding of the chiral phase transition that goes beyond the Hartree-Fock approximation we studied the model by the

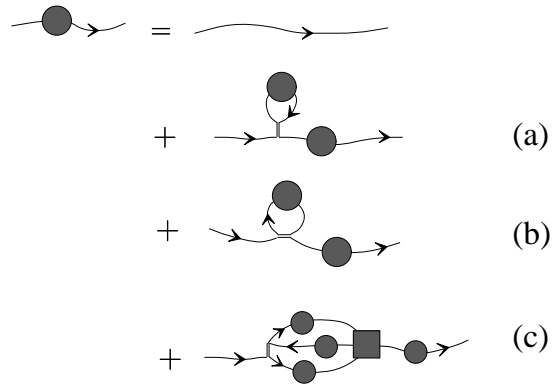

Figure 1. The Schwinger-Dyson equations for the fermion propagator.

Hybrid-Monte Carlo method. Additionally numerical Schwinger-Dyson techniques were applied. To enable a direct comparison we studied the same lattice version of (1) in both cases. Because chiral symmetry is important in the original model we used staggered fermions in our lattice action, since they allow a continuous chiral symmetry on the lattice. The lattice action reads

$$S = \frac{1}{2} \sum_{x,\mu} \eta_\mu(x) \left[ \bar{\chi}(x)\chi(x+\hat{\mu}) - \bar{\chi}(x+\hat{\mu})\chi(x) \right]$$
$$+ m_0 \sum_x \bar{\chi}(x)\chi(x)$$
$$- g_0 \sum_{x,\mu} \bar{\chi}(x)\chi(x)\bar{\chi}(x+\hat{\mu})\chi(x+\hat{\mu}) \quad (2)$$

where $\eta_\mu(x) = (-1)^{x_1+\cdots+x_{\mu-1}}$, $\eta_1(x) = 1$. For

---

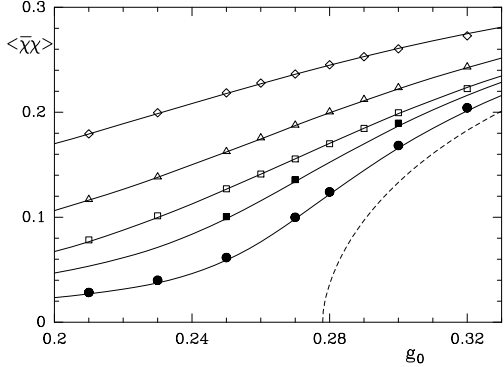

Figure 2. The Monte Carlo data for the chiral condensate and our fit. Symbol shape denotes bare mass (diamonds 0.09, triangles 0.04, squares 0.02 and circles 0.01). $12^4$ lattices are black points, $8^3 \times 16$ white.

the Monte Carlo treatment the four fermion term was rewritten in a quadratic form by introducing an auxiliary field $\theta_\mu(x) \in [0, 2\pi]$:

$$
\begin{aligned}
S_{\text{int}} =\ & \sqrt{g_0} \sum_{x,\mu} \eta_\mu(x) \left[ \bar{\chi}(x) e^{i\theta_\mu(x)} \chi(x + \hat{\mu}) \right. \\
& \left. - \bar{\chi}(x + \hat{\mu}) e^{-i\theta_\mu(x)} \chi(x) \right] .
\end{aligned}
\tag{3}
$$

In Fig. 1 we represent graphically the Schwinger-Dyson equations for the fermion pro-pagator. The equations as presented there are exact, but must be truncated to give a numerically tractable system of equations. The first approximation (known as the gap equation) keeps only the term labelled (a). This is the leading term in the usual $1/N$ expansion. The next approximation is to keep (a) and (b). We refer to this system as the $O(g_0)$ equations. Finally we keep all three terms but approximate the full four-point function by the bare four-point vertex. This approximation is referred to as the $O(g_0^2)$ equations.

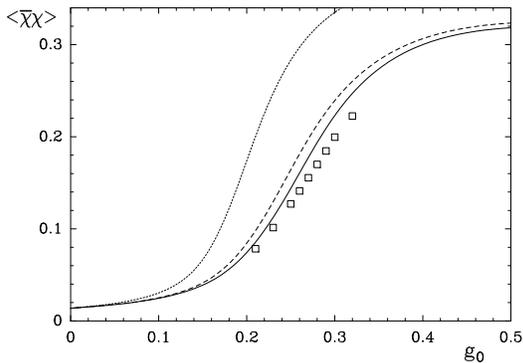

Figure 3. A comparison between the Monte Carlo data for the chiral condensate and the three truncations of the Schwinger-Dyson equations. The lattice size is $8^3 \times 16$, $m_0 = 0.02$.

The two methods have different strengths and weaknesses. The Monte Carlo yields results without any approximations, but can only be used on relatively small lattices. Comparing the results of the methods we were able to check that the approximations made in the Schwinger-Dyson equations are reasonable. Knowing this we could confidently use the Schwinger-Dyson technique to go to large and even infinite lattices.

We measured the chiral condensate $\langle \bar{\chi}\chi \rangle$,

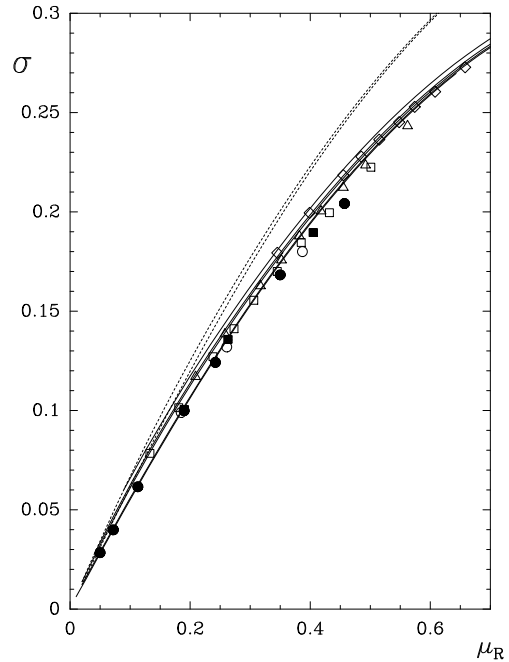

Figure 4. A plot of the chiral condensate $\sigma$ against the renormalised fermion mass. Symbols as in Fig. 2, lines as in Fig. 3.

the renormalised fermion mass $\mu_R$, energies of fermion-antifermion composite states and the pion decay constant $f_\pi$. With these results at hand we considered renormalisation group flows.

## 2. THE CHIRAL CONDENSATE AND CRITICAL COUPLING

First we looked at the chiral condensate in order to map out the phase diagram. Using a fit *ansatz* that was based on a modified gap equation we determined the critical coupling to be $g_c \approx 0.278$. The data and fit are shown in Fig. 2.

In Fig. 3 we compare Monte Carlo results for the chiral condensate with the predictions of three different truncations of the Schwinger-Dyson equations. The agreement improves as more terms are included, and the final curve is only slightly shifted from the true results.

The fermion mass $\mu_R$ behaves similarly to the chiral condensate. A plot (Fig. 4) of $\langle \bar{\chi}\chi \rangle$ vs. $\mu_R$

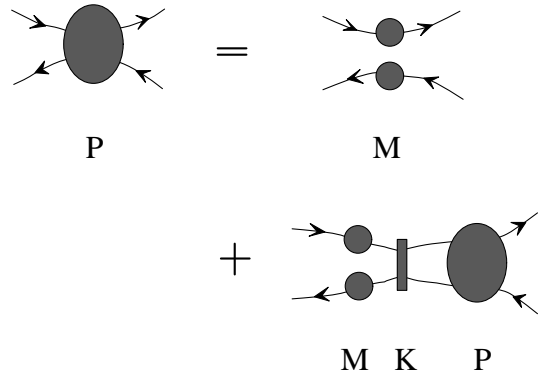

Figure 6. The Schwinger-Dyson equations for the meson propagator $P$.

shows that the data lie on a nearly universal curve. The curves are the predictions of the gap equation and of the $O(g_0^2)$ Schwinger-Dyson equations. We see that the predictions lie in a very narrow band. The differences between the curves are caused by finite size effects and a small residual dependence on bare mass.

As can be seen by comparing the white and black squares in Fig. 2 finite size effects are rather severe in the Nambu–Jona-Lasinio model. It is therefore fortunate that by analysing the Schwinger-Dyson equations we can find a useable description of the finite size effects. For large lattice size $L_s$

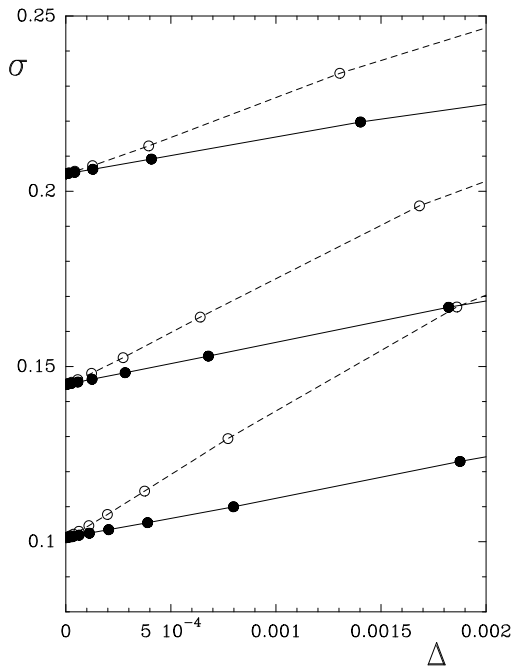

Figure 5. Finite size effects for the chiral condensate on $L_s^4$ (solid points) and $L_s^3 \times 2L_s$ (open points) lattices.

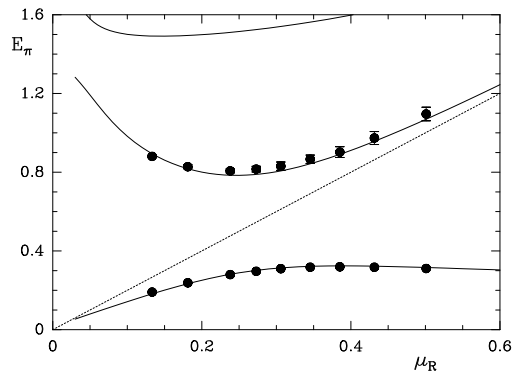

Figure 7. Energy levels for the $\pi$ channel. The dotted line is the threshold $E = 2\mu_R$.

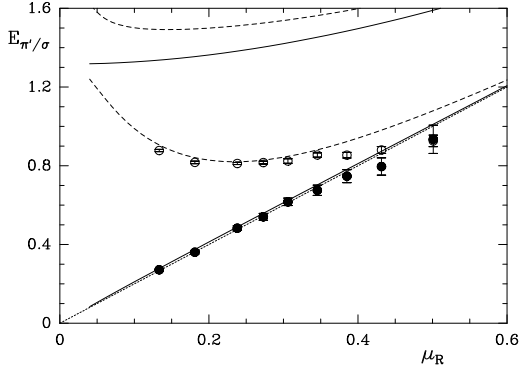

Figure 8. Energy levels in the $\pi'/\sigma$ channel. Solid curves are the $\pi'$ levels (the $s$-wave channel) and the dashed curves are the $\sigma$ levels ($p$-wave).

$$\langle \bar{\chi}\chi \rangle - \langle \bar{\chi}\chi \rangle(\infty) \propto (\mu_R/L_s)^{\frac{3}{2}} \exp(-\mu_R L_s) \equiv \Delta . \quad (4)$$

This equation can be tested by plotting the chiral condensate from differently sized lattices against $\Delta$, as in Fig. 5, (if the equation holds the data should lie on straight lines).

## 3. SPECTROSCOPY

In the meson channels we measured propagators for the local bilinear operators and for wall sources. We compared Monte Carlo results with the solution of the Schwinger-Dyson equations sketched in Fig. 6. The extra information from the wall sources allowed us to find not only the ground state energies but also the first excited state. In Figs. 7 and 8 we compare our Monte Carlo results with the levels predicted (with no adjustable parameters) in the Schwinger-Dyson approach.

To understand the meson sector properly we need to consider the infinite volume limit. (This is particularly important for the levels above the threshold $E = 2\mu_R$.) Therefore in Fig. 9 we have plotted the $\pi$ ground state energies on increasingly large lattices ($8^3 \times \infty$, $12^3 \times \infty$, $20^3 \times \infty$ and $52^3 \times \infty$). For the largest lattice (thin solid lines) we have also shown the excited states. Also shown are Monte Carlo data from $12^4$ and $8^3 \times 16$ lattices. On an infinite lattice the ex-cited states form a continuum. In the symmetric phase there is a resonance in this continuum, which we have plotted as a thick line. On the $52^3 \times \infty$ lattice there is a sequence of avoided level crossings These are signatures of the resonance on a finite lattice. For the pseudoscalar meson ($\pi$) we found a light bound state in the broken phase. This mass vanishes like $\sqrt{m_0}$ as expected from chiral perturbation theory. In the symmetric phase the $\pi$ turned out to be a resonance. The energy levels of the scalar meson ($\sigma$) obtained by the Schwinger-Dyson method are also shown in Fig. 10. The $\sigma$ is a resonance (bold line) if $g_0 \approx g_c$ and in the symmetric phase. Deep in the broken phase it becomes a very weakly bound state. In the whole broken phase the $\sigma$ mass is about $2\mu_R$.

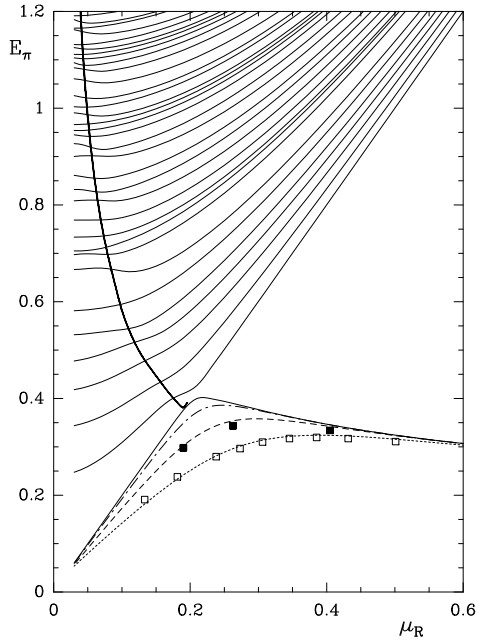

Figure 9. Pion energy levels in increasing volumes.

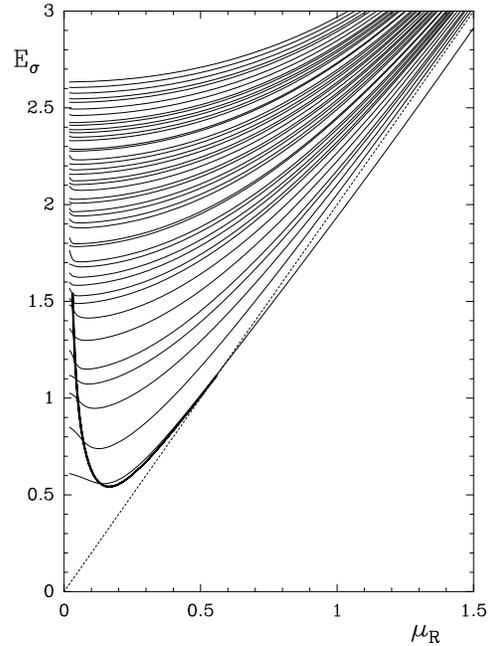

Figure 10. The energy levels of the scalar meson. ($20^3 \times \infty$ lattice, $m_0 = 0.02$).

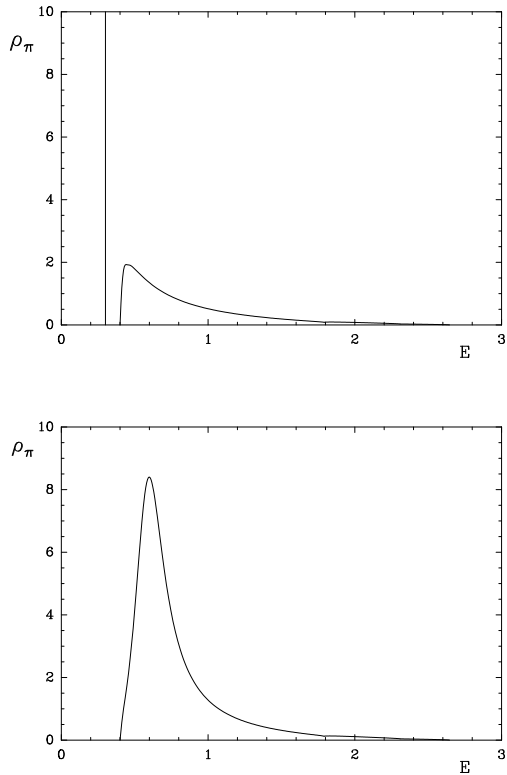

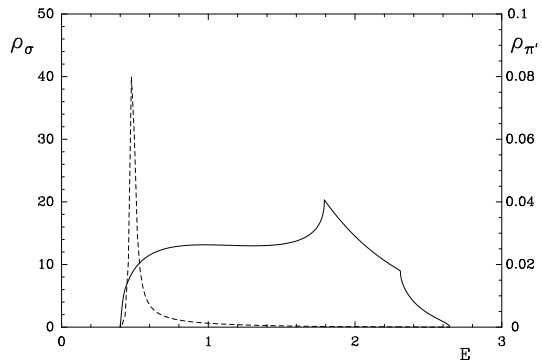

Figure 12. The spectral function for the scalar particle (dashed curve) and the $\pi'$ (solid curve). The scalar is a narrow resonance near the threshold. (Note that the two curves are shown at different scales.)

Figure 11. The pion spectral function. In the first picture, from the broken phase, the pion is a true bound state and the spectral function has a delta function at $E = m_\pi$. In the second diagram the pion is a resonance, able to decay into a fermion-antifermion pair.

Using the Schwinger-Dyson equations we can also investigate the limit of infinite lattice volume, where the closely spaced energy levels lying above the threshold go over into a continuum. In this case we can define a spectral function $\rho$. In Figs. 11 and 12 we show spectral functions for the pion and scalar channels. We see delta functions when a true bound state exists, and resonances when a state is heavy enough to decay into two fermions. We found the masses and widths of the resonances by locating poles on the second Riemann sheet in complex energy [5].

The vector meson ($\rho$) also turned out to be a resonance. In contrast to the $\pi$ and $\sigma$ which become massless at the critical point the $\rho$ does not become light. Its mass scales with the cutoff when the critical point is approached.

## 4. RENORMALISATION GROUP TRAJECTORIES

Using the results from spectroscopy we obtained renormalisation group flows, i.e., contour plots of dimensionless ratios of physical quantities in the plane of the bare parameters $m_0$ and $g_0$. In particular we looked at the ratios $m_\pi/\mu_R$ and $\mu_R/f_\pi$, Fig. 13. We found that the lines of constant $m_\pi/\mu_R$ flow into the critical point while the lines of constant $\mu_R/f_\pi$ end on the $m_0$ axis. The two sets of lines cross everywhere in the area around the critical point. Hence there cannot be lines of constant physics in this region which implies the absence of renormalisability. Some caution has to be exercised in interpreting a flow diagram measured on a finite lattice. As we have seen the ground state energy in the pion channel is a true bound state in the broken phase but in the symmetric phase it corresponds to an unbound fermion antifermion pair with an energy strongly influenced by the lattice volume. We have therefore repeated the renormalisation group trajectory on infinite lattices using the Schwinger-Dyson equations. The resulting flow diagrams are shown in Figs. 14 and 15. The flow lines cross, so there are no lines of constant physics.

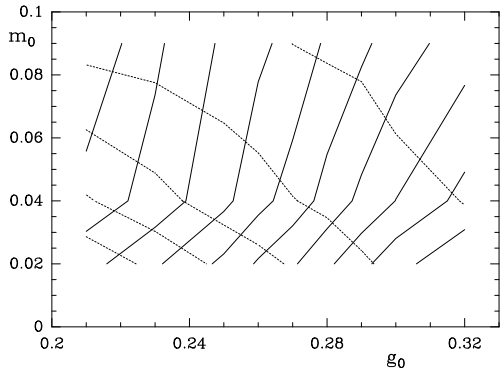

Figure 13. A comparison between the flows of constant $m_\pi/\mu_R$ (solid lines) and constant $\mu_R/f_\pi$ (dotted lines) on $8^3 \times 16$ lattices.

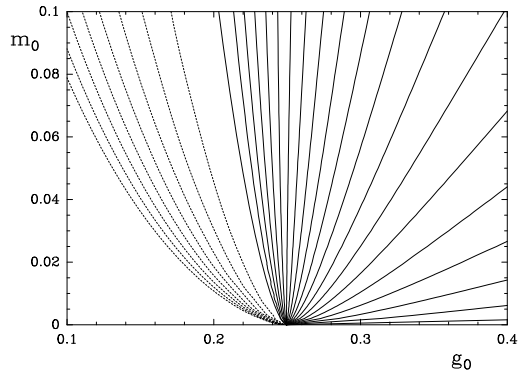

Figure 14. The $m_\pi/\mu_R$ flow on an infinite lattice. (Dotted line resonance, solid bound state.)

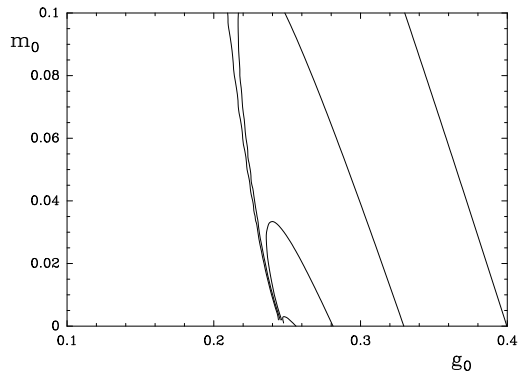

Figure 15. The $\mu_R/f_\pi$ flow on an infinite lattice.

## 5. ACKNOWLEDGEMENTS

This work was supported in part by the Deutsche Forschungsgemeinschaft. The numerical calculations were performed on the Fujitsu VP 2400/40 at the RRZN Hannover and the Crays at the ZIB Berlin and the RZ Kiel. We thank all these institutions for their support.